\documentclass[aps,pra,showpacs,twocolumn,superscriptaddress]{revtex4}

\usepackage{amsmath,amssymb,amsfonts,epsfig}

\begin{document}

\title{Heralded Entanglement of Arbitrary Degree in Remote Qubits}

\author{U. Schilling}
\affiliation{Institut f\"{u}r Optik, Information und Photonik, Max-Planck Forschungsgruppe, Universit\"{a}t Erlangen-N\"{u}rnberg, 91058 Erlangen, Germany}

\author{C. Thiel}
\affiliation{Institut f\"{u}r Optik, Information und Photonik, Max-Planck Forschungsgruppe, Universit\"{a}t Erlangen-N\"{u}rnberg, 91058 Erlangen, Germany}

\author{E. Solano}
\affiliation{Departamento de Qu\'{i}mica F\'{i}sica, Universidad del Pa\'{i}s Vasco -- Euskal Herriko Unibertsitatea, Apdo. 644, 48080 Bilbao, Spain}

\author{T. Bastin}
\affiliation{Institut de Physique Nucl\'{e}aire, Atomique et de Spectroscopie, Universit\'{e} de Li\`{e}ge, 4000 Li\`{e}ge, Belgium}

\author{J. von Zanthier}
\affiliation{Institut f\"{u}r Optik, Information und Photonik, Max-Planck Forschungsgruppe, Universit\"{a}t Erlangen-N\"{u}rnberg, 91058 Erlangen, Germany}

\date{\today}

\begin{abstract}
Incoherent scattering of photons off two remote atoms with a $\Lambda$-level structure is used as a basic Young-type interferometer to herald long-lived entanglement of an arbitrary degree. The degree of entanglement, as measured by the concurrence, is found to be tunable by two easily accessible experimental parameters. An estimate of the variation of the degree of entanglement due to uncertainties in an experimental realization is given.
\end{abstract}


\maketitle

Quantum interference~\cite{Mandel:1995} and entanglement~\cite{Nielsen:2000} are two of the most stunning consequences of quantum mechanics. Although these phenomena are usually studied separately, both have quantum parallelism as a common origin: Quantum interference deal with the coherent superposition of multiple quantum paths, typically for a single system, while entanglement is inherent to the nonseparable character of linear superpositions in the multipartite case. This common origin leads to the possibility of constructing tight links between both phenomena. For example, Jakob and Bergou~\cite{Bergou:2003,Bergou:2007} derived a relation between the entanglement of two qubits and the visibility of the interference pattern generated by one of the qubits in a Ramsey-type interferometer. Another way to link these two properties is to couple interfering quantum paths to remote physical systems. With this ansatz, Scholak \emph{et al.}~\cite{Scholak:2008} showed that the interference pattern of a single photon probing two spatially separated atomic systems can {\it witness} their mutual entanglement. 

With a similar approach, by detecting the interference pattern of scattered photons, it is also possible to \emph{create} entanglement among the particles~\cite{Bose:1999,Cabrillo:1999,Duan:2001,Skornia:2001,Simon:2003,Duan:2003,Moehring:2007,Thiel:2007,Bastin:2007}.
Hereby, the atoms may be separated by arbitrary distances, as there is no need for a particle interaction. This should be contrasted with other schemes entangling massive particles, which require some kind of interaction, be it Coulomb-like~\cite{Turchette:1998,Solano:1999,Haeffner:2005,Leibfried:2005} or mediated by photons~\cite{Hagley:1997,Zheng:2000,Polzik:2001,Osnaghi:2001,Matsukevich:2006}. Furthermore, at variance with photon entanglement, usually achieved  by parametric down conversion~\cite{Kwiat:1995,Kiesel:2007}, the entanglement of electronic ground states of atoms can be preserved over long periods of time~\cite{Haeffner:2005,Leibfried:2005}. This may prove to be useful for diverse applications in quantum communication and quantum computation~\cite{Nielsen:2000}, where long-lived entanglement plays a crucial role.

In this paper, we present a proposal involving a simple scheme to operationally tune the amount of long-lived entanglement present in two remote atomic qubits. 
We will demonstrate that our scheme allows to create heralded entangled states, where the degree of entanglement between the atomic qubits can be tuned at will. The exact value is determined by adjusting two easily accessible experimental parameters, namely the position of two photodetectors and the relative orientation of two polarizers.

\begin{figure}[t]
\centering
\includegraphics[scale=0.8]{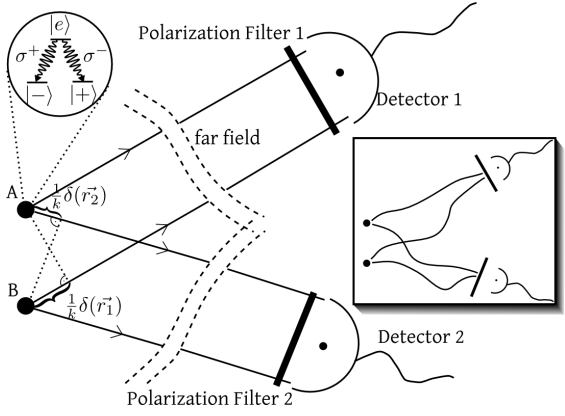}
\caption{Scheme of two atoms with internal $\Lambda$-level structures using two detectors in the far field with polarization filters in front to register the photons emitted by the atoms. The inset shows the same configuration using optical fibers.}
\label{fig_lambda-level_setup}
\end{figure}

The proposed setup is based on a Young-type interferometer realized by two localized atoms~\cite{Eichmann:1993} with an internal $\Lambda$-level structure (see Fig.~\ref{fig_lambda-level_setup}). The atoms, representing the double-slit of the interferometer, are excited by a laser pulse and subsequently scatter photons in their deexcitation process. Without loss of generality, we will assume that the upper state (denoted $|e\rangle$) decays to the lower states (denoted $|\pm\rangle$) by emitting a $\sigma^{\mp}$ polarized photon, respectively. These photons are registered in the far field with photon detectors, which are additionally equipped with polarization filters in front of them. The far-field detection is a simple method to erase the which-way information of the photons propagating from the atoms to the detector. The atoms are projected by the measurement of the two photons into a given state, depending on the position of the detectors and the detected polarizations~\cite{Thiel:2007,Bastin:2007}. For an arbitrary two-qubit pure state $|\psi\rangle = a|++\rangle +b |+-\rangle + c |-+ \rangle + d |--\rangle$ the concurrence reads 
\begin{eqnarray}
\mathcal C = |\langle \tilde{\psi} | \psi \rangle | = 2 |ad -bc| .
\label{concurrence}
\end{eqnarray}
where $|\tilde{\psi}\rangle = (\sigma_{y,A} \otimes \sigma_{y,B}) |\psi\rangle$, with $\sigma_{y,X}$ the usual $\sigma_{y}$ Pauli matrix of the qubit formed by the two lower states of atom $X$ ($X = A,B$)~\cite{Wootters:1998}.

Omitting proportionality factors, the detection of a photon scattered off two $\Lambda$-level atoms $A$ and $B$, with a detector $D_i$ at position $\vec{r}_i$ behind a polarization filter aligned along $\vec{\varepsilon}_i$, is described by the projection operator~\cite{Agarwal:2002}
\begin{equation}
\hat{D}_i = \hat{D}_i(\vec{r}_i) = \hspace{-3pt}\sum_{m = \pm} (\vec{\varepsilon}_i \cdot \vec{d}_{me})\left( |m\rangle_A\langle e| + e^{-i \delta(\vec{r}_i)} |m\rangle_B \langle e|\right) ,
\label{general_detection_operator}
\end{equation}
where the sum runs over the two ground states $|\pm\rangle$. Here, $\vec{d}_{\pm e}$ is the dipole moment of the transition $|e\rangle \rightarrow |\pm \rangle$ and the phase difference $\delta (\vec{r}_i)$ is given by
\begin{equation}
\delta (\vec{r}_i) = k (\vec{R}_B-\vec{R}_A) \cdot \vec{e}(\vec{r}_i) ,
\end{equation}
where $k$ is the wavenumber of the detected photon, $\vec{R}_{A,B}$ is the position of the respective atom, and $\vec{e}(\vec{r}_i)$ is the unit vector pointing from the atoms towards the detector $D_i$ at $\vec{r}_i$. Hereby, the far-field detection ensures that $\vec{e}(\vec{r}_i)$ is identical for both atoms.

The far-field detection scheme provides for the loss of which-way information of the scattered photons. The same can be accomplished by using optical fibers guiding the photons from the atoms to the detectors~\cite{Volz:2006,Moehring:2007,Eschner:2008}. In this case, the phase difference $\delta(\vec{r}_i)$ is given by
\begin{equation}
\delta(\vec{r_i})= k\left(w_B(\vec{r_i})-w_A(\vec{r_i})\right) ,
\end{equation}
where $w_{A,B}(\vec{r}_i)$ is the optical path length from the respective atom to the detector $D_i$ at position $\vec{r}_i$ via the corresponding optical fiber (cf. Fig.~\ref{fig_lambda-level_setup}). In this configuration the atoms can be separated by arbitrary distances, i.e., they are truly remote.

By applying the operator $\hat{D}_1$ and $\hat{D}_2$ to the initial double excited state of the two atoms, $|\psi^{(i)}_{\Lambda}\rangle = | e e \rangle$, we find the normalized atomic state after the detection of two photons to be
\begin{multline}
|\psi^{(f)}_{\Lambda}\rangle = \frac{\hat{D}_1 \hat{D}_2 |\psi^{(i)}_{\Lambda}\rangle}{\sqrt{\langle \psi^{(i)}_{\Lambda} | \hat{D}_2^{\dagger}\hat{D}_1^{\dagger} \hat{D}_1 \hat{D}_2 |\psi^{(i)}_{\Lambda}\rangle}} =\\
\zeta\left[ \left(1 + e^{- i \delta_{21}}\right) \left(\varepsilon_{2-} \varepsilon_{1-}|++\rangle + \varepsilon_{2+} \varepsilon_{1+} |--\rangle \right)\right.\\
+\left( \varepsilon_{2+} \varepsilon_{1-} + e^{- i \delta_{21}} \varepsilon_{2-} \varepsilon_{1+} \right) |-+\rangle \\
+ \left. \left( e^{- i \delta_{21}} \varepsilon_{2+} \varepsilon_{1-} + \varepsilon_{2-} \varepsilon_{1+} \right) |+-\rangle\right] .
\label{lambda_after_two_photon_emission}
\end{multline}

Here, the abbreviation $\varepsilon_{i\pm} = \vec{\varepsilon}_i \cdot \vec{d}_{\mp e}$ is used, where without loss of generality we assume $|\varepsilon_{i+}|^2+|\varepsilon_{i-}|^2=1$, $\delta_{21}$ is given by the phase difference
\begin{equation}
\delta_{21} = \delta(\vec{r}_2) - \delta(\vec{r}_1),
\end{equation}
depending on the two detector positions $\vec{r}_1$ and $\vec{r}_2$, and $\zeta$ is a normalization factor. 

Using Eq.~(\ref{concurrence}), the concurrence of the pure state Eq. (\ref{lambda_after_two_photon_emission}) can be explicitly calculated. One obtains
\begin{equation}
\mathcal C(\delta_{21},\mathcal V_{12}) = \frac{\left| \varepsilon_{2+} \varepsilon_{1-} - \varepsilon_{2-} \varepsilon_{1+} \right|^2}{1+\left| \vec{\varepsilon}_2 \cdot \vec{\varepsilon}^{\,\, *}_1\right|^2 \cos \delta_{21}} = \frac{1 - \mathcal V_{12}}{1 + \mathcal V_{12} \cos \delta_{21}},
\label{lambda_concurrence}
\end{equation}
where the parameter $\mathcal V_{12}$ is given by
\begin{equation}
\mathcal V_{12}=|\vec{\varepsilon}_2 \cdot \vec{\varepsilon}_1^{\,\, *}|^2 .
\label{g2_visibility}
\end{equation}
According to Eq.\,(\ref{lambda_concurrence}), the long-lived entanglement generated between the ground states of the two $\Lambda$-level atoms only depends on the relative phase $\delta_{21}$ and on the relative orientation of the two polarization filters $\mathcal V_{12}$ (see Fig.~2)~\cite{footnote}. In order to obtain a certain amount of entanglement between the two atoms, both parameters have to be tuned to suitable values and the excitation of the atoms has to be repeated until both detectors register a photon. By postselection we then know that the atomic pair contains exactly the desired amount of entanglement as described by Eq.\,(\ref{lambda_concurrence}).
\begin{figure}[ht]
\begin{center}
\includegraphics[scale=0.3]{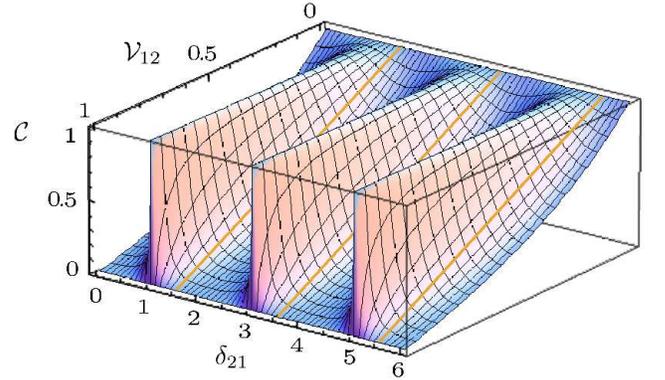}
\caption{The concurrence as a function of the phase $\delta_{21}$ (scaled in multiples of $\pi$) and the parameter $\mathcal V_{12}$. The thick orange lines mark constant $\delta_{21}=(n+1/2)\pi$, where the dependence of $\mathcal C$ on $\mathcal V_{12}$ becomes linear.}
\label{concurrence_fig}
\end{center}
\end{figure}
Taking a look at the extremal values of $\mathcal C$ with respect to $\delta_{21}$, we obtain
\begin{eqnarray}
\mathcal C_{\text{min}} & = & \frac{1-\mathcal V_{12}}{1+\mathcal V_{12}} \hspace{1.5em}\text{if}\hspace{1.5em} \cos \delta_{21} = 1 \, , \nonumber \\
\mathcal C_{\text{max}} & = & 1  \hspace{4.5em}\text{if}\hspace{1.5em} \cos \delta_{21} = -1 \, .
\label{delta_extremum_concurrence}
\end{eqnarray}
These expressions show that, depending on the value of $\mathcal V_{12}$, \emph{any} amount of concurrence between $\frac{1-\mathcal V_{12}}{1+\mathcal V_{12}}$ and 1 can be achieved. In particular, by choosing $\delta_{21}$ to be an odd multiple of $\pi$, it is always possible to generate a state with maximal (unit) concurrence, independent of the explicit value of $\mathcal V_{12}$, i.e., independent of the relative orientation of the two polarization filters (see Fig.~\ref{concurrence_fig}).

The extrema of the concurrence with respect to $\mathcal V_{12}$ are given by
\begin{eqnarray}
\mathcal C = 1 \hspace{1em}\text{ for \hspace{0.6em} all \, $\mathcal V_{12}$,\hspace{1em} if }\cos\delta_{21} = -1\nonumber\\
\left.
\begin{matrix}
\mathcal C_{\text{min}} = 0 \hspace{1em}\text{ for }\hspace{1em} \mathcal V_{12}=1,\\
\mathcal C_{\text{max}} = 1 \hspace{1em}\text{ for }\hspace{1em} \mathcal V_{12}=0,
\end{matrix}
\right\}\text{ if }\cos\delta_{21} \neq -1.
\label{v_extremum_concurrence}
\end{eqnarray}
Thus, if the phase difference is not fixed to an odd multiple of $\pi$, it is always possible to use $\mathcal V_{12}$ as a single parameter to tune the concurrence to any desired value. In particular, by choosing $\delta_{21}$ an odd multiple of $\pi/2$, we find in Eq.~(\ref{lambda_concurrence}) a \emph{linear} relation between the concurrence and the parameter $\mathcal V_{12}$. In this case, when linear polarizers are used, by keeping one of them fixed and turning the other by a relative angle $\alpha$, we are able to implement a fully tunable concurrence ($0 < \mathcal C < 1$)
\begin{equation}
\mathcal C = 1-\mathcal V_{12} = \sin^2 \alpha ,
\label{malus_law}
\end{equation}
yielding an analog to the Malus' Law \cite{Hecht:2002}. In its classical version, it says that the intensity of the {\it same} light beam passing consecutively through two linear polarizers is proportional to the square of the cosine of the relative angle between the polarizers. Here, we find that the concurrence, a measure characterizing the entanglement of two qubits, behaves in a similar way. Even though each of the two indistinguishable photons passes a {\it different} polarizer, the degree of entanglement generated between the atoms upon detection of the photons is determined by the relative angle between the two polarizers. This result can be seen as an operational implementation of a tunable measure of entanglement between matter qubits following a simple and intuitive law of classical optics.

Note that the parameter $\mathcal V_{12}$ intervenes also in the second order correlation function $G^{(2)}(\delta_{21})$, which is proportional to the measured signal. The second order correlation function reads~\cite{Agarwal:2002}
\begin{eqnarray}
G^{(2)}(\delta_{21}) &=& \langle \psi^{(i)}_{\Lambda} | \hat{D}_2^{\dagger} \hat{D}_1^{\dagger} \hat{D}_1 \hat{D}_2 |\psi^{(i)}_{\Lambda} \rangle\nonumber\\
&=&2 \left( 1 + \mathcal V_{12} \cos \delta_{21}\right) ,
\label{lamda_g2}
\end{eqnarray}
In this expression $\mathcal V_{12}$  appears as the {\em visibility} of the $G^{(2)}(\delta_{21})$-function, revealing again the close relationship between quantum interference and entanglement.

In the following, we will give an estimate of the variation of the concurrence due to experimental uncertainties (see also \cite{Thiel:2007,Bastin:2007}). The probability to detect a scattered photon is proportional to the solid angle subtended by the detector divided by $4\pi$. By extending the detection area, the detection probability will increase, though the accumulated phase will be less well defined. Thus, there is a trade-off between the count rate of the scattered photons and the error in the concurrence generated in the final state. For estimating errors, we will assume identical rectangular detectors. Let $\alpha_D$ be the azimuthal angular extension of each detector in direction of $\theta_i$, with $\theta_i$ the azimuth angle between $\vec{e}(\vec{r_i})$ and the axis connecting the two atoms, and $\varphi_D$ the polar angle subtended by each detector perpendicular to the plane of $\alpha_D$. Then, for small $\alpha_D$, the probability to detect a randomly emitted photon with one of the two detectors can be approximated to
\begin{equation}
P(\alpha_D,\varphi_D) = \frac{\alpha_D\varphi_D}{4\pi}.
\end{equation}
The count rate $R$ of two-photon detection events is thus given by
\begin{equation}
R = 2 r \cdot P(\alpha_D,\varphi_D)^2 \cdot G^{(2)}(\delta_{21}) ,
\label{count_rate}
\end{equation}
where $r$ is the repetition rate of the experiment. A factor of 2 appears since either detector, $D_1$ or $D_2$, might register the first photon. The count rate is thus maximal if the condition for constructive interference of the second order correlation function $G^{(2)}(\delta_{21})$ is fulfilled, i.e., if $\delta_{21}$ is an even multiple of $\pi$.

The uncertainty in the concurrence $\Delta \mathcal C$ is defined by the difference in the concurrence of the density matrix of the state $\rho^{(g)}$ actually generated and the pure target state $\rho^{(f)}=|\psi^{(f)}_{\Lambda}\rangle \langle \psi^{(f)}_{\Lambda}|$:
\begin{equation}
\Delta \mathcal C = |\mathcal C(\rho^{(g)})-\mathcal C(\rho^{(f)})|.
\end{equation}
$\Delta \mathcal C$ is essentially determined by the uncertainty in the orientation of the polarization filters $\Delta \mathcal V$ and the uncertainty in the phase $\Delta \delta_{21}$. With current experimental technology, $\Delta \mathcal V$ can be suppressed to the order of $10^{-10}$~\cite{Tamburini:2008}. Thus, $\Delta \mathcal V$ is negligible compared to the uncertainty imposed by the phase and will be neglected in the following.

The uncertainty in the phase $\Delta\delta_{21}$ is governed by two contributions: the solid angle subtended by the detector (determined by $\alpha_D$ and $\varphi_D$) and the finite confinement $\mu$ of the atoms in the trap. To calculate $\rho^{(g)}$, we have to integrate over the whole relevant parameter space:
\begin{equation}
\int_{A,\mu} w(\delta_{21})\;|\psi^{(f)}_{\Lambda}\rangle\langle\psi^{(f)}_{\Lambda}|(\delta_{21})\,{\rm d}A\,{\rm d}\mu
\end{equation}
where $w(\delta_{21})$ is a weight factor determined by the geometry of the setup and normalization. To minimize the deviation from the desired final state, we have to minimize $\varphi_D$ and $\alpha_D$, while $\theta_i$ should be close to $\frac{\pi}{2}$. However, $\varphi_D$ and $\alpha_D$ are bound from below by the requirement of an acceptable count rate $R$. In addition, there is a lower boundary to $\Delta\mathcal C$ due to the finite confinement of the atoms. 

For realistic experimental parameters, $d=5 \, \mu\text{m}$, $\mu = 10 \, \text{nm}$, $\alpha_D = 5 \, \text{mrad}$, $\varphi_D=\frac{\pi}{6}$, $\theta \approx \frac{\pi}{2}$, and photons of wavelength $\lambda=650 \, \text{nm}$, this results in $\Delta \mathcal C_{\text{max}}< 0.025$ for all $\delta_{21} \in [-\frac{\pi}{2},\frac{\pi}{2}]$ and all $\mathcal V$. Within this parameter range, the fidelity of the final state $\rho^{(g)}$ always remains above 95\,\%, while for a repetition rate $r$ of a few Mhz the count rate amounts to a few events per second. These estimates include a detector efficiency of about 30\,\% and a dark count rate of up to a few 100\,Hz.

The protocol presented here is capable of producing heralded entangled states with a high fidelity. The count rate, on the other hand, is relatively low in the analysed case. Modifications in the setup concerning the detector shape and the number of detectors are possible, as well as the use of fibers or cavities to increase the detection probability of the scattered photons without curtailing the fidelity. These suggested modifications do not change the principal results of this paper, but they might contribute to a better implementation of the presented basic ideas.

In conclusion, we have shown that with a simple and realistic setup, it is possible to create heralded entanglement of any degree between two remote atoms with a $\Lambda$-type level structure. As the atoms are entangled by projective measurements requiring no atomic interaction, the atomic distances in a given experiment are arbitrary. In particular, instead of using a far-field measurement to erase the which-way information of photons, the use of optical fibers could provide a more practical approach to reach similar goals. We expect that our results inspire and stimulate further research in operational and realistic methods for the generation and measure of entanglement in different experimental contexts.

U.S. thanks the Elite Network of Bavaria for financial support. E.S. thanks funding from Ikerbasque Foundation, EU EuroSQIP, and UPV-EHU Grant GIU07/40. C.T. and J.v.Z. gratefully acknowledge financial support by the Staedtler foundation. The authors thank G. S. Agarwal for fruitful discussions.


\begin{thebibliography}{10}

\bibitem{Mandel:1995}
L. Mandel and E. Wolf, {\it Optical Coherence and Quantum Optics} (Cambridge University Press, Cambridge, England, 1995).

\bibitem{Nielsen:2000}
M. Nielsen and I. Chuang, {\it Quantum Computation and Quantum Information} (Cambridge University Press, Cambridge, England, 2000).

\bibitem{Bergou:2003}
M. Jakob and J.~A. Bergou, arXiv: quant-ph/0302075.

\bibitem{Bergou:2007}
M. Jakob and J.~A. Bergou, Phys. Rev. A {\bf 76}, 052107 (2007).

\bibitem{Scholak:2008}
T. Scholak, F. Mintert, and C.~A. M\"{u}ller, Europhys. Lett. {\bf 83}, 60006 (2008).

\bibitem{Bose:1999}
S.~Bose, P.~L. Knight, M.~B. Plenio, and V.~Vedral, Phys. Rev. Lett. {\bf 83}, 5158 (1999).

\bibitem{Cabrillo:1999}
C.~Cabrillo, J.~I. Cirac, P.~Garc\'ia-Fern\'andez, and P.~Zol\-ler, Phys. Rev. A {\bf 59}, 1025 (1999).

\bibitem{Duan:2001}
L.-M. Duan, M.~D. Lukin, J.~I. Cirac, and P.~Zoller, Nature {\bf 414}, 413 (2001).

\bibitem{Skornia:2001}
C.~Skornia, J.~von Zanthier, G.~S. Agarwal, E.~Werner, and H.~Walther, Phys. Rev. A {\bf 64}, 063801 (2001).

\bibitem{Simon:2003}
C. Simon and W.~T.~M. Irvine, Phys. Rev. Lett. {\bf 91}, 110405 (2003).

\bibitem{Duan:2003}
L.-M. Duan and H.~J. Kimble, Phys. Rev. Lett. {\bf 90}, 253601 (2003).

\bibitem{Moehring:2007}
D.~L. Moehring, P.~Maunz, S.~Olmschenk, K.~C. Younge, D.~N. Matsukevich, L.-M.
Duan, and C.~Monroe, Nature {\bf 449}, 68 (2007).

\bibitem{Thiel:2007}
C. Thiel, J. von Zanthier, T. Bastin, E. Solano, and G.~S. Agarwal, Phys. Rev. Lett. {\bf 99}, 193602 (2007).

\bibitem{Bastin:2007}
T. Bastin, C. Thiel, J. von Zanthier, L. Lamata, E. Solano, and G.~S. Agarwal, Phys. Rev. Lett. {\bf 102}, 053601 (2009). 

\bibitem{Turchette:1998}
Q.~A. Turchette, C.~S. Wood, B.~E. King, C.~J. Myatt, D.~Leibfried, W.~M. Itano,  C.~Monroe, and D.~J. Wineland, Phys. Rev. Lett. {\bf 81}, 3631 (1998).

\bibitem{Solano:1999}
E. Solano, R. L. de Matos Filho, and N. Zagury, Phys. Rev. A {\bf 59}, R2539 (1999); {\bf 61}, 029903(E) (2000).

\bibitem{Haeffner:2005}
H.~H\"{a}ffner, W.~H\"{a}nsel, C.F. Roos, J.~Benhelm, D.~{Chek-al-kar},
M.~Chwalla, T.~K\"{o}rber, U.D. Rapol, M.~Riebe, P.O. Schmidt, C.~Becher,
O.~G\"{u}hne, W.~D\"{u}r, and R.~Blatt, Nature {\bf 438}, 643 (2005).

\bibitem{Leibfried:2005}
D.~Leibfried, E.~Knill, S.~Seidelin, J.~Britton, R.B. Blake\-stad, J.~Chiaverini, D.B. Hume, W.M. Itano, J.D. Jost, C.~Langer, R.Ozeri, R.~Reichle, and D.J. Wineland, Nature {\bf 438}, 639 (2005).

\bibitem{Hagley:1997}
E.~Hagley, X.~Ma\^itre, G.~Nogues, C.~Wunderlich, M.~Brune, J.M. Raimond, and
S.~Haroche, Phys. Rev. Lett. {\bf 79}, 1 (1997).

\bibitem{Zheng:2000}
S.-B. Zheng and G.-C. Guo, Phys. Rev. Lett. {\bf 85}, 2392 (2000).

\bibitem{Polzik:2001}
B. Julsgaard, A. Kozhekin, and E. S. Polzik,  Nature {\bf 413}, 400 (2001).

\bibitem{Osnaghi:2001}
S.~Osnaghi, P.~Bertet, A.~Auffeves, P.~Maioli, M.~Brune, J.~M. Raimond, and
S.~Haroche, Phys. Rev. Lett. {\bf 87}, 037902 (2001).

\bibitem{Matsukevich:2006}
D.~N. Matsukevich, T.~Chaneli\`{e}re, S.~D. Jenkins, S.-Y. Lan, T.~A.~B. Kennedy,
and A.~Kuzmich, Phys. Rev. Lett. {\bf 96}, 030405 (2006).

\bibitem{Kwiat:1995}
P. G. Kwiat, K.Mattle, H. Weinfurter, A. Zeilinger, A. V. Sergienko, and Y. Shih,
Phys. Rev. Lett., {\bf 75}, 4337 (1995).

\bibitem{Kiesel:2007}
N. Kiesel, C. Schmid, G. T\'oth, E. Solano, and H. Weinfurter,
Phys. Rev. Lett. {\bf 98}, 063604 (2007).

\bibitem{Eichmann:1993}
U. Eichmann, J.~C. Bergquist, J.~J. Bollinger, J.~M. Gilligan, W.~M. Itano, D.~J. Wineland, M.~G. Raizen, Phys. Rev. Lett. {\bf 70}, 2359 (1993).

\bibitem{Wootters:1998}
W.~K. Wootters, Phys. Rev. Lett. {\bf 80}, 2245 (1998).

\bibitem{Agarwal:2002}
G.~S. Agarwal, J. von Zanthier, C. Skornia, and H. Walther, Phys. Rev. A {\bf 65}, 053826 (2002).

\bibitem{Volz:2006}
J. Volz, M. Weber, D. Schlenk, W. Rosenfeld, J. Vrana, K. Saucke, C. Kurtsiefer, and H. Weinfurter, Phys. Rev. Lett. {\bf 96}, 030404 (2006).

\bibitem{Eschner:2008}
S.~Gerber, D.~Rotter, M.~Hennrich, R.~Blatt, F.~Rohde, C.~Schuck, M.~Almendros,
 R.~Gehr, F.~Dubin, and J.~Eschner, arXiv: quant-ph/0810.1847.
 
\bibitem{footnote}
Note that Eq.~(\ref{lambda_concurrence}) only holds if $\mathcal V_{12}$ and $\cos \delta_{21}$ are not simultaneously equal to 1 and $-1$, respectively, since in this case no two-photon signal is detected (cf. Eq.~(\ref{lamda_g2})) so that the projection described in Eq.~(\ref{lambda_after_two_photon_emission}) does not occur.

\bibitem{Hecht:2002}
Eugene Hecht, {\em Optics} (Addison Wesley, San Francisco, California, USA, 2002).

\bibitem{Tamburini:2008}
F. Tamburini, B.~A. Bassett, and C. Ungarelli, Phys. Rev. A {\bf 78}, 012114 (2008).
\end{thebibliography}
\end{document}